\input epsf
\documentstyle[12pt,epsf]{article}
\newlength{\pubnumber} 

\catcode`\@=11
\@addtoreset{equation}{section}

\def\section{\@startsection{section}{1}{\z@}{3.5ex plus 1ex minus .2ex}
 {2.3ex plus .2ex}{\large\bf}}
\def\subsection{\@startsection{subsection}{2}{\z@}{2.3ex plus .2ex}
 {2.3ex plus .2ex}{\bf}}

\font\bigbf=cmssbx10 scaled\magstep2
\font\it=cmti10 at 12pt
\font\ss=cmss10 at 12pt

 at 10truept

\ss 					%% default font
\begin{document}

\begin{titlepage}
\samepage{
\setcounter{page}{1}
%\rightline{Preliminary DRAFT}
\rightline{UFIFT--HEP--96--17}
\rightline{\tt hep-ph/9608420}
\rightline{August 1996}
\vfill
\begin{center}
 {\Large \bf  Cosmological and phenomenological implications \\
		of Wilsonian matter in \\
		realistic superstring
               derived models\footnote{talk presented at String 96,
		July 15--20 1996, Santa Barbara, CA}}
\vfill
\vfill
 {\large Alon E. Faraggi\footnote{
        E-mail address: faraggi@phys.ufl.edu}\\}
\vspace{.12in}
 {\it   Institute for Fundamental Theory, Department of Physics, \\
        University of Florida, Gainesville, FL 32611,
        USA\\}
\end{center}
\vfill
\begin{abstract}
  {\rm Superstring phenomenology aims at achieving two goals.
The first is to reproduce the observed physics of the Standard Model.
The second is
to identify experimental signatures of superstring unification
which, if observed, will provide further evidence for the
validity of superstring theory. I discuss
such potential signatures of superstring unification.
I propose that proton lifetime constraints imply that
the Standard Model gauge group must be obtained directly at the string level.
In this case the unifying gauge group, for example $SO(10)$, is
broken to the Standard Model gauge group by ``Wilson lines''.
The symmetry breaking by ``Wilson line'' has important implications.
It gives rise to exotic massless states which cannot fit into multiplets
of the original unifying gauge group. This is an important feature
because it results in conservation laws which forbid the interaction
of the exotic ``Wilsonian'' states with the Standard Model states.
The ``Wilsonian'' matter states then have  important phenomenological
implications. I discuss two such implications: exotic ``Wilsonian''
states as dark matter candidates and  ``Wilsonian'' matter as
the messenger sector in gauge mediated dynamical SUSY breaking scenarios.}
\end{abstract}
\smallskip}
\end{titlepage}

\setcounter{footnote}{0}

% ========================= DEFINITIONS ===================================
\def\beq{\begin{equation}}
\def\eeq{\end{equation}}
\def\beqn{\begin{eqnarray}}
\def\eeqn{\end{eqnarray}}

\def\ie{{\it i.e.}}
\def\eg{{\it e.g.}}
\def\half{{\textstyle{1\over 2}}}
\def\third{{\textstyle {1\over3}}}
\def\quarter{{\textstyle {1\over4}}}
\def\m{{\tt -}}
\def\p{{\tt +}}

\def\slash#1{#1\hskip-6pt/\hskip6pt}
\def\slk{\slash{k}}
\def\GeV{\,{\rm GeV}}
\def\TeV{\,{\rm TeV}}
\def\y{\,{\rm y}}
\def\SM{Standard-Model }
\def\SUSY{supersymmetry }
\def\SSSM{supersymmetric standard model}
\def\vev#1{\left\langle #1\right\rangle}
\def\l{\langle}
\def\r{\rangle}

\def\Htw{{\tilde H}}
\def\chibar{{\overline{\chi}}}
\def\qbar{{\overline{q}}}
\def\ibar{{\overline{\imath}}}
\def\jbar{{\overline{\jmath}}}
\def\Hbar{{\overline{H}}}
\def\Qbar{{\overline{Q}}}
\def\abar{{\overline{a}}}
\def\alphabar{{\overline{\alpha}}}
\def\betabar{{\overline{\beta}}}
\def\tautwo{{ \tau_2 }}
\def\thetatwo{{ \vartheta_2 }}
\def\thetathree{{ \vartheta_3 }}
\def\thetafour{{ \vartheta_4 }}
\def\ttwo{{\vartheta_2}}
\def\tthree{{\vartheta_3}}
\def\tfour{{\vartheta_4}}
\def\ti{{\vartheta_i}}
\def\tj{{\vartheta_j}}
\def\tk{{\vartheta_k}}
\def\calF{{\cal F}}
\def\smallmatrix#1#2#3#4{{ {{#1}~{#2}\choose{#3}~{#4}} }}
\def\ab{{\alpha\beta}}
\def\Minv{{ (M^{-1}_\ab)_{ij} }}
\def\bone{{\bf 1}}
\def\ii{{(i)}}
\def\V{{\bf V}}
\def\b{{\bf b}}
\def\N{{\bf N}}
\def\t#1#2{{ \Theta\left\lbrack \matrix{ {#1}\cr {#2}\cr }\right\rbrack }}
\def\C#1#2{{ C\left\lbrack \matrix{ {#1}\cr {#2}\cr }\right\rbrack }}
\def\tp#1#2{{ \Theta'\left\lbrack \matrix{ {#1}\cr {#2}\cr }\right\rbrack }}
\def\tpp#1#2{{ \Theta''\left\lbrack \matrix{ {#1}\cr {#2}\cr }\right\rbrack }}
\def\l{\langle}
\def\r{\rangle}

%================== BLACKBOARD BOLD CHARACTERS ==============================

\def\inbar{\,\vrule height1.5ex width.4pt depth0pt}

\def\IC{\relax\hbox{$\inbar\kern-.3em{\rm C}$}}
\def\IQ{\relax\hbox{$\inbar\kern-.3em{\rm Q}$}}
\def\IR{\relax{\rm I\kern-.18em R}}
 \font\cmss=cmss10 \font\cmsss=cmss10 at 7pt
\def\IZ{\relax\ifmmode\mathchoice
 {\hbox{\cmss Z\kern-.4em Z}}{\hbox{\cmss Z\kern-.4em Z}}
 {\lower.9pt\hbox{\cmsss Z\kern-.4em Z}}
 {\lower1.2pt\hbox{\cmsss Z\kern-.4em Z}}\else{\cmss Z\kern-.4em Z}\fi}

%========================================================================
%          MACROS FOR REFERENCES
%========================================================================
\def\AEF{A.E. Faraggi}
\def\NPB#1#2#3{{\it Nucl.\ Phys.}\/ {\bf B#1} (19#2) #3}
\def\PLB#1#2#3{{\it Phys.\ Lett.}\/ {\bf B#1} (19#2) #3}
\def\PRD#1#2#3{{\it Phys.\ Rev.}\/ {\bf D#1} (19#2) #3}
\def\PRL#1#2#3{{\it Phys.\ Rev.\ Lett.}\/ {\bf #1} (19#2) #3}
\def\PRT#1#2#3{{\it Phys.\ Rep.}\/ {\bf#1} (19#2) #3}
\def\MODA#1#2#3{{\it Mod.\ Phys.\ Lett.}\/ {\bf A#1} (19#2) #3}
\def\IJMP#1#2#3{{\it Int.\ J.\ Mod.\ Phys.}\/ {\bf A#1} (19#2) #3}
\def\nuvc#1#2#3{{\it Nuovo Cimento}\/ {\bf #1A} (#2) #3}
\def\etal{{\it et al\/}}

%==============================================================================
\hyphenation{su-per-sym-met-ric non-su-per-sym-met-ric}
\hyphenation{space-time-super-sym-met-ric}
\hyphenation{mod-u-lar mod-u-lar--in-var-i-ant}
%==============================================================================

%============================== SECTION 1 ============================

\setcounter{footnote}{0}
Superstring phenomenology aims at achieving two goals.
The first is to reproduce the observed physics. The second
is to identify possible experimental signature of superstring
unification which may provide further evidence for its validity.
The first task is highly nontrivial and indeed only a handful
of string models can claim to be potentially realistic. Indeed
the number of constraints is large and satisfying all in one
string model is an almost impossible challenge. A model which
satisfies all of the constraints of the observed low energy
physics, is likely to be more than an accident. Such a model,
or class of models, will then serve as the laboratory for the
search for exotic predictions of superstring unification. It
will also serve as a laboratory in which we can address the
important question of how the string vacuum is selected.

\smallskip
A few of the constraints that a realistic model of unification
must satisfy are listed below.

\centerline{{$\underline{{\hbox{~~~~~~~~~~~~~~~~~~~~~~~~~~~~~~~~~~~~~}}}$}}

{}~~~~~$1.$ Gauge group ~$\longrightarrow$~ $SU(3)\times SU(2)\times U(1)_Y$

{}~~~~~$2.$ Contains three generations

{}~~~~~$3.$ Proton stable ~~~~~~~~($\tau_{\rm P}>10^{30+}$ years)

{}~~~~~$4.$ N=1 supersymmetry~~~~~~~~(or N=0)

{}~~~~~$5.$ Contains Higgs doublets $\oplus$ potentially realistic
Yukawa couplings

{}~~~~~$6.$ Agreement with $\underline{\sin^2\theta_W}$ and
$\underline{\alpha_s}$ at $M_Z$ (+ other observables).

{}~~~~~$7.$ Light left--handed neutrinos

{}~~~~~~~~~~~~~$~8.$ $SU(2)\times U(1)$ breaking

{}~~~~~~~~~~~~~$~9.$ SUSY breaking

{}~~~~~~~~~~~~~$10.$ No flavor changing neutral currents

{}~~~~~~~~~~~~~$11.$ No strong CP violation

{}~~~~~~~~~~~~~$12.$ Exist family mixing and weak CP violation

{}~~~~~$13.$ +~~ {\bf ...}

{}~~~~~$14.$ +~~~~~~~~~~~~~~~~{\bigbf{GRAVITY}}

\centerline{{$\underline{{\hbox{~~~~~~~~~~~~~~~~~~~~~~~~~~~~~~~~~~~~~}}}$}}

\smallskip
The first question that we must ask is whether it is possible
to construct a model which satisfies all of those criteria, or
possibly a class of models which can accommodate most of these
constraints. To date the most developed theory that can
consistently unify gravity with the gauge interactions is string
theory \cite{Sreviews}. While alternatives may exist, it makes sense
at this stage to try to use string theory to construct a model which
satisfies the above requirements. Even if eventually string theory
turns out not to be the fundamental theory of nature, a model which
satisfies all of above constraints is likely to arise as an effective
model from the true fundamental theory.

Semi--realistic string models were constructed by using various
methods \cite{otherrsm,revamp,fny,rffm,eu,top,slm}.
Of the above requirements the most difficult
to satisfy in a realistic model is the constraint of proton
stability. The reason is that supersymmetric models are in
general plagued with dimension four and five operators which
give rise to fast proton decay \cite{SUSYreviews}. In supersymmetric
point field theory models such operators may be avoided by
postulating the existence of some symmetries which forbid the
dangerous operators. However, in string models we don't have this
luxury. The desired symmetries either exist in the models or they
do not. Often one can find that at some points in the
moduli space of specific models the dangerous operators
vanish due to some accidental global symmetry. This is
not quite satisfactory for two reasons. First, in general the
isolated points in the moduli space will not accommodate some other
constraints, like potentially non--vanishing Yukawa couplings which
can give rise to fermion masses. Second, in general global
symmetries are badly broken in string models and the dangerous
operators can be induced from non--vanishing nonrenormalizable terms
which effectively reproduce the dangerous dimension four and five
operators. The longevity of the proton lifetime
imposes that such nonrenormalizable operators must
be suppressed to very high orders.
For these reasons, we would like the desired symmetry, which
suppresses the proton decay operators, to be a robust
gauge symmetry or a local discrete symmetry.

The problem of proton stability suggests that
the allowed gauge groups at the string scale
are restricted to very few choices.
First, it is desirable to avoid giving
a VEV of the order of the GUT or Planck scale to
the right handed neutrino. Otherwise, effective dimension
four operators may be induced. Second, if the symmetry
is broken at the string level to $SO(6)\times SO(4)$
or directly to $SU(3)\times SU(2)\times U(1)^2$ then
there is a superstring doublet--triplet splitting
mechanism in which the triplets are projected from
the massless spectrum by the GSO projections
while the doublets remain in the light spectrum \cite{ps}.
Therefore, if the symmetry at the string scale is broken to
$SO(6)\times SO(4)$ or directly to the Standard Model
gauge group the problems with proton decay can be avoided
in a robust way. Thus, due to proton lifetime constraints,
the prefered symmetries at the string level are
$SO(6)\times SO(4)$ or $SU(3)\times SU(2)\times U(1)^2$.

Another very restrictive constraint on realistic superstring
models is the requirement of agreement with $\sin^2\theta_W$ and
$\alpha_s$ at $M_Z$. This constraint is better known as the string
scale gauge coupling unification problem. If we assume that the
spectrum between the electroweak scale is that of the Minimal
Supersymmetric Standard Model, then it is well known that the
three gauge couplings intersects at a scale which is of the order
$2\times 10^{16}~{\rm GeV}$ \cite{gcumssm}. While the successful
meeting of the couplings in the MSSM is intriguing, it is far from
being well motivated. The MSSM is not a complete theory and clearly
cannot accommodate all of the requirements listed above. Furthermore,
there is nothing special about the spectrum of the MSSM. The assumption
of a minimal spectrum is ad hoc and is not motivated from any
fundamental principles. On the other hand, at tree level, string theory
predicts that the gauge couplings are unified at a scale which is of the
order $4\times 10^{17}~GeV$ \cite{scales}.
Thus, an order of magnitude separates
the string unification scale from the MSSM unification scale.
It would seem that in extrapolation of the gauge parameters
over some fifteen orders of magnitude, a problem involving
a single order of magnitude would have many possible
solutions. Surprisingly, however, the problem is not easily
resolved. In fact, most string models can immediately be discarded
simply because they predict a value for $\sin^2\theta_W$ at the
string scale which is much smaller than the regular GUT prediction.
Thus, these string models predict a $\sin^2\theta(M_Z)$ which
is much smaller than the experimentally observed value
and cannot be adjusted by small correcting effects.
This is the case because in most string models the weak
hypercharge does not have the regular GUT embedding.
Thus, in a realistic string models we would like
the weak hypercharge to have the standard GUT embedding.
One possibility is of course to consider string GUT
models \cite{SGUTS}. However, those will in general
have problems with proton lifetime. Another proposal
\cite{witten} is that nonperturbative string effects
play an important role, and that one of the compactified
dimensions is of the order of the GUT scale. In this case
there may be additional GUT scale color triplets from the massive
string spectrum and one has to ascertain that those do not
cause rapid proton decay.

The class of realistic string models which are constructed
in the free fermionic formulation \cite{FFF} can satisfy both of these
requirements. In fact, to date, these are the only string
models that have been shown to satisfy both of these constraints.
In these models an $SO(10)$ symmetry is broken at the string scale to
$SU(5)\times U(1)$, $SO(6)\times SO(4)$ or
$SU(3)\times SU(2)\times U(1)^2$ and the weak hypercharge
has the standard $SO(10)$ embedding. Also there exist free
fermionic models in which all the dangerous dimension four
and five operators are suppressed to all orders of nonrenormalizable
terms \cite{ps}. This is achieved in models in which the $SO(10)$
symmetry is broken to $SO(6)\times SO(4)$ or
$SU(3)\times SU(2)\times U(1)^2$ and due to the superstring
doublet--triplet splitting mechanism.
Recently, an interesting study of the proton decay problem in the
context of string models was done in Ref. \cite{pati}. It was found
that precisely the sort of symmetries which appear in the free fermionic
models are those needed to prevent proton decay and allow naturally
suppressed neutrino masses.

It has been shown that the free fermionic models can potentially
satisfy many of the other constraints that must be imposed on a
realistic string model \cite{rsm}.
Thus, free fermionic models are candidates
for a realistic string model. After identifying a class of potentially
realistic string models the second question that we must address
is whether there exist some generic signature of these models
which, if observed, will provide further support for their validity.

In this talk I discuss such possible exotic signatures.
In the free fermionic models, and in string models in general,
one starts with a larger symmetry group which is subsequently broken
to some intermediate unifying symmetry by the GSO projection.
For example in the free fermionic models we start with a
$SO(44)$ gauge group which is broken to $SO(10)\times SO(6)^3\times E_8$.
In the free fermionic models the breaking is achieved by defining
boundary condition basis vectors for the world--sheet fermions
which satisfy some string consistency constraints \cite{FFF}. These
intermediate gauge symmetry is broken further by means
of additional boundary condition basis vectors.
In particular the $SO(10)$ symmetry, in which the Standard Model
is embedded is broken to one of its subgroups, $SU(5)\times U(1)$,
$SO(6)\times SO(4)$ or $SU(3)\times SU(2)\times U(1)^2$.
These additional breaking of the $SO(10)$ gauge symmetry
gives rise to massless states that cannot fit into multiplets
of $SO(10)$.
The additional vectors which break the $SO(10)$ symmetry
correspond to Wilson lines in the orbifold models.
I refer to the extra matter which arises from these sectors
as Wilsonian matter. The breaking by Wilson lines may give
rise to local discrete symmetries which forbid the interaction
of the Wilsonian states with the Standard Model states.
This unique stringy phenomena has important implications.
It implies that the exotic Wilsonian matter states are
stable and therefore may be good dark matter candidates.

To understand better the Wilsonian matter phenomena, it is useful
to study the general structure of the realistic free fermionic models.
As mentioned above the basis vectors which define the models are
divided into two parts. The first part consists of the five vectors
of the NAHE set \cite{revamp,slm}.
These basis vectors correspond to $Z_2\times Z_2$
orbifold compactification. The Neveu-Schwarz (NS) sector corresponds
to the untwisted sector of the orbifold models and produces the
generators of the $SO(10)\times SO(6)^3\times E_8$ gauge group.
In addition to the spin two and spin one states the NS sector
also produces three 10 representation of $SO(10)$ and several $SO(10)$
singlets. The three sectors $b_1$, $b_2$ and $b_3$ correspond to the
three twisted sectors of the $Z_2\times Z_2$ orbifold model and produce
48 multiplets in the 16 representation of $SO(10)$.

The correspondence between this class of free fermionic models
and orbifold models can be shown by constructing the
same orbifold compactification which corresponds to some
free fermionic models \cite{ztwo}.
The simplest way to illustrate this is
by adding to the NAHE set a basis vector $X$,
$${X=\{{{\bar\psi}_{1,\cdots,5},{{\bar\eta}^1,{\bar\eta}^2,
{\bar\eta}^3}}\}=1},$$
which extends the
symmetry to $E_6\times SO(4)^3\times E_8$, with $N=1$ supersymmetry
and 24 generations in the 27 representation of $E_6$. The
same model is constructed in the orbifold formulation by
first constructing the toroidally compactified Narain
lattice \cite{narain} which
corresponds to the free fermionic point in the moduli space.
The metric and antisymmetric tensors are give by

\medskip
{$g_{ij}=\left(\matrix{~2&-1& ~0& ~0& ~0& ~0\cr%
-1& ~2&-1& ~0& ~0& ~0\cr%
{}~0&-1& ~2&-1& ~0& ~0\cr%
{}~0& ~0&-1& ~2&-1&-1\cr%
{}~0& ~0& ~0&-1& ~2& ~0\cr%
{}~0& ~0& ~0&-1& ~0& ~2\cr}\right)~~~~~~~~~b_{ij}=\cases{~~g_{ij}&$i>j$\cr%
{}~~0&$i=j$\cr -g_{ij}&$i<j$}$}
\bigskip

where the metric is the Cartan matrix of $SO(12)$. At the point in the
moduli space which corresponds to the free fermionic models, all the
radii of the compactified dimensions are fixed at a specific value.
At this point there are additional massless vectors bosons which
arise due to the wrapping of the string on the compactified dimensions.
At the specific point which correspond to the free fermionic models
the gauge symmetry is enhanced to $SO(12)\times E_8\times E_8$.
If we now mod out by the $Z_2\times Z_2$ discrete symmetry with
standard embedding, we obtain a model with a gauge group and
spectrum which are the same as those which were obtained in the
fermionic construction. This indeed demonstrates that the
basis vectors of the NAHE set correspond to $Z_2\times Z_2$
orbifold models. In the realistic free fermionic models
we replace the vectors $X$, which extends the $SO(10)$ symmetry
to $E_6$, with the vector $2\gamma$,
$${~~~2\gamma~=\{{{\bar\psi}_{1,\cdots,5},{{\bar\eta}^1,{\bar\eta}^2,
{\bar\eta}^3}}},{\bar\phi}_{1,\cdots,4}{\}=1}.$$
With this substitution
the gauge group is $SO(10)\times U(1)^3\times SO(4)^3\times SO(16)$ with
$N=4$ supersymmetry and 24 generation in the 16 of $SO(10)$.
The vectors $b_j+2\gamma$ now give 24 multiplets in the 16 representation
of the hidden $SO(16)$ gauge group. The modification from the
vector $X$ to the vector $2\gamma$ can be regarded as a transition from
a $(2,2)$ model to a $(2,0)$ model, and can also be achieved by redefining
the GSO phases.

At the level of the NAHE set the observable gauge symmetry is
$SO(10)\times SO(6)^3$ and the number of generations is 48,
sixteen from each of the sectors $b_1$, $b_2$ and $b_3$.
The number of generations is reduced to three by adding three
additional boundary condition basis vectors to the NAHE set.
Each one of the sectors $b_1$, $b_2$ and $b_3$ produces exactly
one generation. We observe that the emergence of three generations
in the realistic free fermionic models is deeply rooted in the
underlying $Z_2\times Z_2$ orbifold structure. Each one of the generations
is in the 16 representation of $SO(10)$ decomposed under the
final subgroup of the original $SO(10)$ gauge group.
Thus, we see that the Standard Model spectrum in this models
has an underlying $SO(10)$ unification structure and the weak
hypercharge has the standard $SO(10)$ embedding. This outcome
is very important as these models will share some of the desirable
features of $SO(10)$ GUT unification. The only difference
is that here the $SO(10)$ symmetry is broken at the string theory
level rather than at the field theory level. This distinction has
of course highly nontrivial consequences. In particular, as mentioned
above with regard to the proton lifetime problem.

The breaking of the $SO(10)$ gauge symmetry can be done with the
same boundary condition basis vectors which reduce the number
of generations. The $SO(10)$ symmetry is broken to one of its
subgroups. This is achieved by the assignment of boundary conditions to the
set ${\bar\psi}^{1,\cdots,5}$:
\begin{eqnarray}
{b\{{{\bar\psi}_{1\over2}^{1\cdots5}}\}=
\{{1\over2}~{1\over2}~{1\over2}~{1\over2}~
{1\over2}\}
\Rightarrow SO(10)~\rightarrow~SU(5)\times U(1)},~~~\label{so10to64}\\
{b\{{{\bar\psi}_{1\over2}^{1\cdots5}}\}=\{1 ~~1 ~~1 ~~0 ~~0\}
\Rightarrow SO(10)~\rightarrow~SO(6)\times SO(4)}.~\label{so10to51}
\end{eqnarray}
To break the $SO(10)$ symmetry to $SU(3)\times SU(2)\times
U(1)_C\times U(1)_L$\footnote{$U(1)_C={3\over2}U(1)_{B-L};
U(1)_L=2U(1)_{T_{3_R}}.$}
both steps, (\ref{so10to64}) and (\ref{so10to51}),
are used, in two separate basis vectors. The flavor $SO(6)^3$ symmetries
are also broken by the additional basis vectors to horizontal $U(1)$
symmetries. Similarly, the hidden $E_8$ gauge group is broken as well to one
of its subgroups.
For example in the standard--like models of refs. \cite{eu,top},
$SO(6)^{{1},{2},{3}}~\rightarrow~U(1)^{{1},{2},{3}}\times
U(1)^{{4},{5},{6}}$, and ${E_8~\rightarrow~SU(5)\times SU(3)\times U(1)^2}$
The additional basis vectors which break the $SO(10)$ gauge
symmetry, denoted commonly as $\{\alpha,\beta,\gamma\}$,
correspond to ``Wilson lines'' in the orbifold formulation.

There is one additional sector which is a linear combination
of the basis vectors $\alpha$ and $\beta$ and which does not violate
the underlying $SO(10)$ gauge symmetry. In the standard--like models
this sector is typically the combination $b_1+b_2+\alpha+\beta$.
The states from this sector arise from $SO(10)$ multiplets. Thus,
the NS sector, the three twisted sectors $b_1$, $b_2$ and $b_3$
and the sector $b_1+b_2+\alpha+\beta$ give rise to ``Standard''
massless spectrum. ``Standard'' here means that all the states
from these sectors can fit into ``standard'' multiplets of $SO(10)$,
whereas, as will be shown below, the ``exotic'' Wilsonian matter states
cannot.

The ``standard'' massless spectrum is summarized in the next page.
This spectrum is common to a large class of realistic free fermionic
models which use the NAHE set. The sector $b_1+b_2+\alpha+\beta$ can
give rise to an additional pair of electroweak doublets or to a pair
of color triplets. The important point to note here is that the massless
spectrum from these sectors has the $U(1)$ quantum numbers of the standard
decomposition of the $SO(10)$ gauge group under its $U(1)$ sub-generators.
Thus, all the states from these sectors can fit into standard $SO(10)$
multiplets. Under the assumption that the $U(1)_{Z^\prime}$ combination
which is orthogonal to the weak hypercharge is not broken near the
Planck scale the states from these sectors also completely determine
the qualitative texture of quarks and charged lepton
mass matrices \cite{rsm}.

\bigskip
\centerline{${\underline{{\hbox{The massless spectrum-- ``standard''}}}}$}%
\baselineskip=14pt%
{Three generations from the twisted sectors ~~{$b_1$}, ~~{$b_2$}, ~~{$b_3$}
with horizontal symmetries: }
\baselineskip=16pt%

${b_1:(e_1+u_1)_{{{1\over2},0,0}\atop{{1\over2},0,0}}+(d_1+N_1)_{{{1\over2},0,0}\atop{{{1}\over2},0,0}}+(Q_1)_{{{1\over2},0,0}\atop{-{1\over2},0,0}}+(L_1)_{{{1\over2},0,0}\atop{-{1\over2},0,0}}%
{}~~~\sigma_4~~\sigma_5~~\chi^{12}}$

${b_2:(e_2+u_2)_{{0,{1\over2},0}\atop{0,{1\over2},0}}+(d_2+N_2)_{{0,{1\over2},0}\atop{0,{1\over2},0}}+(Q_2)_{{0,{1\over2},0}\atop{0,-{1\over2},0}}+(L_2)_{{0,{1\over2},0}\atop{0,-{1\over2},0}}%
{}~~~\sigma_2~~\sigma_6~~\chi^{34}}$

${b_3:(e_3+u_3)_{{0,0,{1\over2}}\atop{0,0,{1\over2}}}+(d_3+N_3)_{{0,0,{1\over2}}\atop{0,0,{{1}\over2}}}+(Q_3)_{{0,0,{1\over2}}\atop{0,0,-{1\over2}}}+(L_3)_{{0,0,{1\over2}}\atop{0,0,-{1\over2}}}%
{}~~~\sigma_1~~\sigma_3~~\chi^{56}}$

\bigskip
{}From the Neveu--Schwarz (untwisted sector)
\baselineskip=16pt%

{}~~~~~~~~~~~~~~~~~~~~~~~~~~~~~~~~~~~~~{$~~~ h_{1_{1,0,0}}~~~~~~~~~~%
{\bar h}_{1_{-1,0,0}}~~~~~~~~~~~$}

{}~~~~~~~~~~~~~~~{Higgs doublets}		~~{$~~~ h_{2_{0,1,0}}~~~~~~~~~~{\bar
h}_{2_{0,-1,0}}~~~~~~~~~~~$}

{}~~~~~~~~~~~~~~~~~~~~~~~~~~~~~~~~~~~~~{$~~~ h_{3_{0,0,1}}~~~~~~~~~~%
{\bar h}_{3_{0,0,-1}}~~~~~~~~~~~$}

\baselineskip=8pt
\begin{eqnarray}
SO(10)~{\rm singlets ~with} &U(1)~ {\rm charges}
{}~~{\Phi_{23}~,~{\bar\Phi_{23}}~,~}{\Phi_{12}~,
{}~{\bar\Phi_{12}}~,~}{\Phi_{13}~,~{\bar\Phi_{13}}}\nonumber\\
{}~~~~~~~~~~~~~~~~~~~~~~{\rm and} &U(1)~ {\rm singlets} ~~{\xi_1~,~}
{\xi_2~,~}{\xi_3}~~~~~~~~~~~~~~~~~~~~~~~~~~~~~\nonumber\end{eqnarray}

Sectors $b_j+2\gamma~~j={1},{2},{3}$ %
\baselineskip=16pt%
produce three 16 of the hidden $SO(16)$
decomposed under the final hidden gauge group %
$SU(5)\times SU(3)\times U(1)^2$
\baselineskip=20pt%

\bigskip
{From the sector} ${b_1+b_2+\alpha+\beta}$
\baselineskip=16pt%

{Higgs doublets}~~~~~~~~~~~~~~~~~~~~~~~~~~~~~{$~~ h_{{\alpha\beta}_{-{1\over2},
-{1\over2},0,0,0,0}}~~~~~~~~~
{\bar h}_{{\alpha\beta}_{{1\over2},{1\over2},0,0,0,0}}$}

{$SO(10)$ singlets with $U(1)$ charges
{}~~~~}${\Phi_{\alpha\beta}~,~\Phi_1^\pm~,~\Phi_2^\pm%
{}~,~\Phi_3^\pm}$
%Sectors ${\{b_1,b_2,b_3,\alpha,\beta\}\pm\gamma}$ additional, model dependent,
% vector--like,%
% spectrum. %

\bigskip
\centerline{$\underline{\hbox{All these states
fit into ``standard'' reps. of} ~SO(10)}$}%
\baselineskip=14pt%
\bigskip

As seen above the sectors which are produced by the NAHE set
and the combination $\alpha+\beta$ produce states which fit into
standard $SO(10)$ representations or are $SO(10)$ singlets.
In addition to the ``standard'' spectrum from the sectors above,
there exist in the fermionic models ``exotic'' spectrum which
cannot fit into $SO(10)$ multiplets. These spectrum arises from
sectors which are combinations of the NAHE basis vectors and the
basis vectors $\{\alpha,\beta,\gamma\}$. These combinations produce
the exotic matter in vector--like representations. In general,
unlike the ``standard'' spectrum, the ``exotic'' spectrum is
highly model dependent. We can however classify the exotic matter
according the pattern of the $SO(10)$ symmetry breaking by the specific
sectors. Each of these sectors breaks the $SO(10)$ symmetry to
$SU(5)\times U(1)$, $SO(6)\times SO(4)$ or
$SU(3)\times SU(2)\times U(1)^2$. Thus, in $SU(5)\times U(1)$ models
only one type of the exotic states can appear. Similarly, in
$SO(6)\times SO(4)$ type models another type of exotic states
can appear. Finally, the $SU(3)\times SU(2)\times U(1)^2$ type
models contain both the $SU(5)\times U(1)$ and $SO(6)\times SO(4)$
type states as well as states which are unique to
$SU(3)\times SU(2)\times U(1)^2$ type models.

\baselineskip=30pt%

Using the notation

{}~~~~~~~~~~~~~~
	{$[(SU(3)_C\times U(1)_C); %
(SU(2)_L\times U(1)_L)]_{(~Q_Y~,~Q_{Z^\prime}~,~Q_{\rm e.m.}~)} $}

where $U(1)_Y=1/3 U(1)_C+1/2 U(1)_L$,
$U(1)_{Z^\prime}=U(1)_C-U(1)_L$ and
$U(1)_{\rm e.m.}=T_{3_L}+U(1)_Y$.

The following exotic states appear in the free fermionic models

\smallskip
\baselineskip=16pt%
\centerline{{$\underline{{\hbox{Exotic matter}}}$}}%

$\underline{{SO(6)\times SO(4)}~\hbox{type states}}$

	{$[(3, {1\over2});(1,0)]_{(~1/6~,~1/2~,~1/6~)}~~~~;~~~~%
[({\bar3},-{1\over2});(1,0)]_{(~-1/6~,~-1/2~,~-1/6)}$}

	{$[(1,0);(2,0)]_{(~0~,~0~,~\pm1/2~)}$}

	{$[(1,0);(1,\pm{1})]_{(~\pm1/2~,~\mp1/2~,~\pm1/2~)}$}

	{$[(1,\pm3/2);(1,0)]_{(~\pm1/2~,~\pm1/2~,~\pm1/2~)}$}

\bigskip

$\underline{{SU(5)\times U(1)}\hbox{type states}}$

	{$[(1,\pm3/4);(1,\pm{1/2})]_{(~\pm1/2~,~\pm1/4~,~\pm1/2~)}$}

\bigskip

$\underline{{SU(3)\times SU(2)\times U(1)^2}\hbox{type states}}$

	{$[(3,{1\over4});(1,{1\over2})]_{(~-1/3~,~-1/4~,~-1/3~)}~~~~;~~~~%
[(\bar3,-{1\over4});(1,{1\over2})]_{(~1/3~,~1/4~,~1/3~)}$}

	{$[(1,\pm{3\over4});(2,\pm{1\over2})]_{(~\pm1/2~,~\pm1/4~,~(1,0)~;~(0,-1)~)}$}

	{$[(1,\pm{3\over4});(1,\mp{1\over2})]_{(~0~,~\pm5/4~,~0~)}$}

\baselineskip=14pt%
\centerline{{$\underline{{\hbox{~~~~~~~~~~~~~~~~~~~~~~}}}$}}%
\smallskip

Thus the exotic states which appear in the $SU(5)\times U(1)$ and
$SO(6)\times SO(4)$
type sectors are fractionally charged with electric charges $\pm1/2$.
The $SU(3)\times SU(2)\times U(1)^2$ type sectors give rise
to matter states which have the ``standard'' charges under the
Standard Model gauge group but carry ``fractional''
charges under the $U(1)_{Z^\prime}$ gauge symmetry.

To examine the phenomenology of these exotic states we
have to study their interactions. The interaction terms in the
superpotential are obtained by calculating the
correlators between vertex operators \cite{kln}
$$\langle{V_1^fV_2^fV_3^b\cdot\cdot\cdot\cdot V_N^b\rangle}$$
The non--vanishing correlators must be invariant under all the
symmetries and the string selection rules.

The free fermionic models contain an anomalous $U(1)$ gauge symmetry.
The anomalous $U(1)$ generates a Fayet--Iliopoulos D--term which
breaks supersymmetry at the Planck scale \cite{dsw}.
The anomaly is canceled by
assigning VEVs to some Standard Model singlets in the massless
string spectrum. In general, these singlets are
charged also with respect to other $U(1)$ symmetries
of a given string model. Therefore, requiring that all the D-terms
vanish imposes a nontrivial set of constraints on the allowed VEVs.
In addition we require that all the F--terms vanish which imposes
that the superpotential and all of its derivatives vanish to all
order of nonrenormalizable terms. Some of the higher order
nonrenormalizable terms become effective renormalizable operators
due to the VEVs which are used to cancel the anomalous
$U(1)$ D--term equation.

The implications of the ``Wilsonian'' matter states were studied in
detail in several examples of realistic free fermionic standard--like
models. The first example is the model of ref. \cite{gcu}. In this model
the $SO(10)$ gauge group is broken to $SU(3)\times SU(2)\times U(1)^2$.
It contains three generations and electroweak Higgs doublet which can
generate phenomenologically realistic fermion mass spectrum. The top
Yukawa has been calculated from cubic order term in the superpotential.
The bottom quark and tau lepton Yukawa were calculated from quartic
order terms and there exist mass and generation mixing terms for
the lighter two generations from higher order terms. In this model
there are no dimension four operators which can
mediate proton decay due to a custodial gauge symmetry \cite{custodial}.
All the ``standard'' color triplets are projected out from the
massless spectrum. As a result all the dimension five operators
vanish as well. Thus, in this model proton decay may only arise from
the massive string spectrum. This model contains exotic ``Wilsonian''
color triplets and electroweak doublets in vector--like representations.
String--scale gauge coupling unification in this model is compatible
with low energy data, provided that the ``Wilsonian'' color triplets
and electroweak doublets exist at the appropriate mass thresholds.

The exotic ``Wilsonian'' matter states appear in the free fermionic
models in vector--like representations, and obtain mass terms from
cubic level or nonrenormalizable terms in the superpotential.

The first example of exotic ``Wilsonian'' matter states are
the states with fractional electric charge which appear in
$SU(5)\times U(1)$ and $SO(6)\times SO(4)$ type sectors. As
there are strong limits on the existence of free fractionally
charged states, this states must be very massive, confined
into integrally charged states, or diluted away. For example,
in the flipped $SU(5)$ model \cite{revamp,elnfc} all the
fractionally charged states transform under a non--Abelian
hidden gauge group and are therefore confined.
In ref. \cite{fc} it was shown that all the fractionally charged
states in the model of ref. \cite{fny} get mass terms at the cubic
level of the superpotential by giving VEVs to four Standard Model
singlets in the spectrum of that model along a Flat F and D direction.
Thus, in this model all the fractionally charged states receive Planck
scale mass and decouple from the light spectrum. Finally, the
fractionally charged states may of course be diluted away. In which
case their abundance today may be too small to be detected experimentally.

In addition to the fractionally charged states, the free fermionic
standard--like models contain ``Wilsonian'' states which
carry the regular charges under the Standard Model gauge group
but carry ``fractional'' charges under the $U(1)_{Z^\prime}$
symmetry, which exist in $SO(10)$. These states can be color
triplets, electroweak doublets, or Standard Model singlets
and may be good candidates for dark matter. The first example
of such states are the color triplets. The existence of color
triplets at intermediate energy scale is motivated from the
problem of gauge coupling unification. The analysis of ref.
\cite{df} showed that heavy string thresholds \cite{moduli},
light SUSY thresholds, intermediate gauge structure,
hypercharge normalization \cite{hynormalization}
do not resolve the problem. The analysis there suggests that the
only way to resolve the problem is the existence of additional intermediate
matter thresholds, beyond the spectrum of the MSSM \cite{gaillard}.
The additional
color triplet thresholds are in general much lighter than the additional
electroweak doublets. The extra color triplets and electroweak doublets
do appear in some string models \cite{gcu}.
Specifically, in the model of ref.
\cite{gcu} the extra states appear from ``Wilsonian'' sectors and
are therefore candidates for ``Wilsonian'' dark matter. Due to its
role in the string gauge coupling unification problem, this type
of color triplet is referred to as the uniton \cite{ccf}.

The uniton has the ``standard'' charges under the standard model
gauge group and ``fractional'' $U(1)_{Z^\prime}$ charge.
The possible interactions of the uniton with the Standard Model states
are:

\smallskip
{$LQ{\bar D},~u_L^ce_L^cD,~QQD,~u_L^cd_L^c{\bar D},~d_L^cN_L^cD,$

$QDh$

${\bar D}{\bar D}u_L^c$}
\smallskip

The terms in the first line above are of the form $b_ib_jD\phi^{n-3}$.
Where $b_i$ and $b_j$ are states from the sectors $b_1$, $b_2$ and $b_3$,
$D$ is the uniton and $\phi^{n-3}$ is a string of Standard Model singlets
which insures that a given term satisfies all the string selection rules.
because of the ``fractional'' $U(1)_{Z^\prime}$ charge of the uniton
all the terms above break $U(1)_{Z^\prime}$. Therefore, the string
$\phi^{n-3}$ should contain ``fractional'' $U(1)_{Z^\prime}$ charge.
In this model the only Standard Model singlets with ``fractional''
$U(1)_{Z^\prime}$ charge are triplets of $SU(3)_H$.
The same is true for the interaction terms from the
second and third lines. Therefore, if
we assume that the $SU(3)_H$ gauge group is unbroken, then all the
interaction terms with the Standard Model states vanish to all orders
of nonrenormalizable terms. In this case the uniton is stable.

The uniton is stable and may be a candidate for the dark matter.
Under the Standard Model gauge group the uniton has the same
charges as a down quark and is strongly interacting. It forms
bound meson states with the regular up and down quarks: $U^\pm$,
$U^0$. An important question is which of the states, the charged
or the neutral, is the lighter state. The mass difference is
determined by the current mass difference and the interaction mass
difference. The mass difference between the charged and neutral
state cannot be calculated reliably because of nonperturbative
color interactions. In ref. \cite{ccf} we studied the mass splitting
by using heavy quark effective theory as well as potential models. We
argued that with our present understanding of QCD and our present
knowledge of the experimental data, it impossible to conclude which
of the mesons is the lighter one. Thus, we argued that at present
there exist a window in the parameter space for which
$M_{U^\pm}>M_{U^0}$. In this case the uniton may be a good
dark matter candidate.

\begin{figure}[htb]
\centerline{\epsfbox{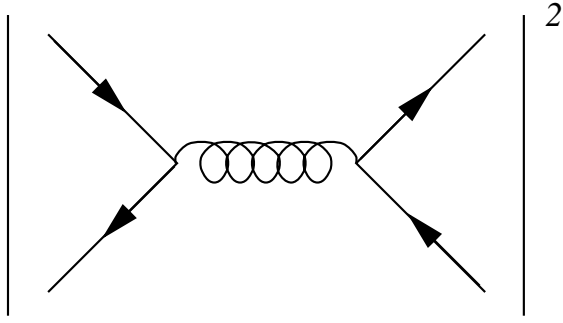}}
\caption{$Q\bar{Q}\longrightarrow q\bar{q}$ annihilation}
\label{aquarks}
\end{figure}

\begin{figure}[htb]
\centerline{\epsfbox{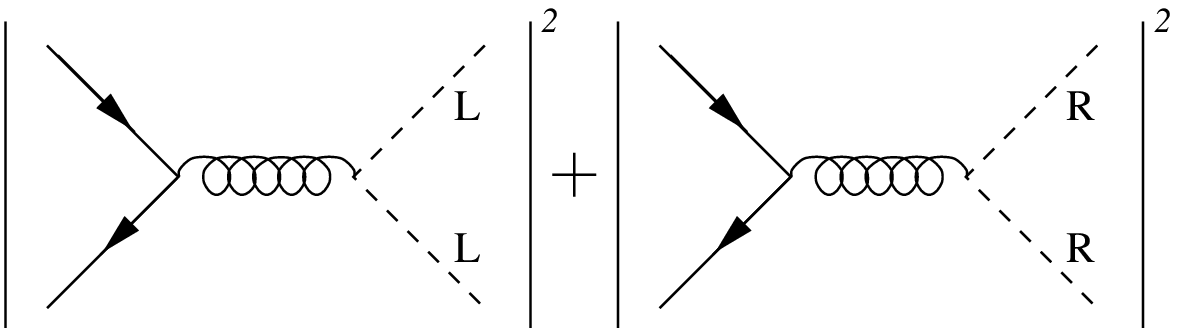}}
\caption{$Q\bar{Q}\longrightarrow \tilde{q}\tilde{q}^*$ annihilation}
\label{asquarks}
\end{figure}

\begin{figure}[htb]
\centerline{\epsfbox {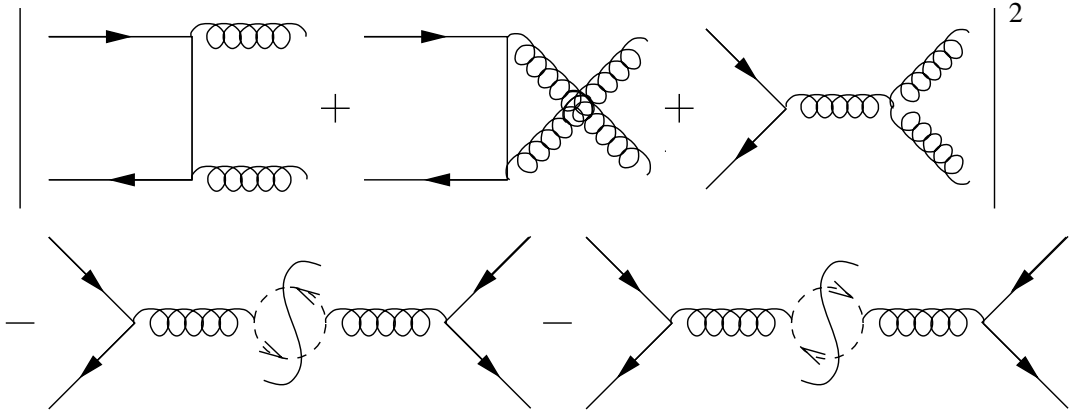}}
\caption{$Q\bar{Q}\longrightarrow gg$ annihilation}
\label{agluons}
\end{figure}

\begin{figure}[htb]
\centerline{\epsfbox{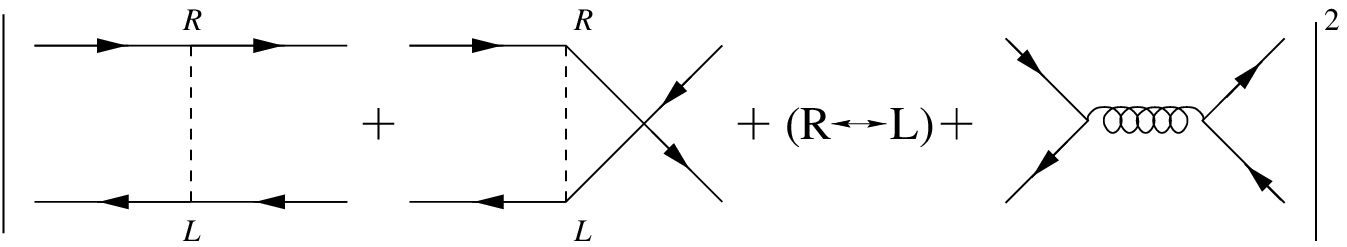}}
\caption{$Q\bar{Q}\longrightarrow \tilde{g}\tilde{g}$ annihilation}
\label{QQgtgt}
\end{figure}

I now discuss the cosmological and astrophysical bounds on the
uniton \cite{ccf}. The uniton is a strongly interacting particle.
In the early universe it remains in thermal equilibrium
until it becomes non--relativistic.
The uniton decouples from the thermal bath when its annihilation rate
falls behind the expansion rate of the universe. In the non--relativistic
limit, $T/M<1$, the uniton annihilation rate
is given by
$$\Gamma =
 <\sigma|v|> n_{eq}
\simeq\frac{\pi N \alpha_s^2 }{M^2 } n_{eq},$$
where $M$ is the mass of the uniton, $\alpha_s$ is the strong coupling at
decoupling, $n_{eq}$ is the number density of the uniton at
equilibrium,
$$n_{eq}=g_{\rm eff}\left({{mT}\over{2\pi}}\right)^{3/2}\exp(-m/T) \ %
{}~~~~({\rm non-relativistic}),$$ and
$N$ is a summation over all the available annihilation
channels and is given by
$$N=\sum_f{a_f}$$
The amplitudes $a_f$ are obtained by calculating the annihilation cross
section of the uniton to all the strongly interacting particles,
which include the six flavors of quarks (fig.\ \ref{aquarks})
and squarks (fig.\ \ref{asquarks})
and the gluons (fig.\ \ref{agluons})
and
the gluinos (fig.\ \ref{QQgtgt}).
The final states are taken to be massless, yielding
$a={4/3}$ for quarks; $a=14/27$ for gluons;
$a=2/3$ for squarks and $a=64/27$ for gluinos.
In the calculation of the cross section one has to:
``properly'' include the ghost contribution to
remove the unphysical polarization contribution in the
annihilation into gluons (fig. \ref{agluons}.

In the expanding universe,
the evolution equation of the particle density
in comoving volume is
$$\frac{dY}{dx} = -\lambda x^{-2} (Y^2-Y^2_{eq}). $$
Here $Y=n/s$, $x\equiv M/T$ and
\begin{eqnarray}
\lambda&=&\left.\frac{x<\sigma|v|> s}{H}\right|_{x=1}
=0.83 N\alpha_s^2\frac{g_*s}{\sqrt{g_*}}\frac{m_{pl}}{ M}.
\label{lam}
\end{eqnarray}
Here the entropy $s$ is $(2\pi^2/45) g_{*s} m^3 x^{-3}$.
The decoupling condition $dY/dx \simeq 0$ gives \cite{KTB}
$$x_{dec} = \ln [(2+c)\lambda ac] -
\frac{1}{2}\ln\{\ln [(2+c)\lambda ac]\},$$
where $ a= 0.145(g/g_{*s})$ and $c$ is
$Y(T_{dec})/Y_{eq}(T_{dec})$, which is of
order one.
We approximately estimate the decoupling temperature to be of the form
$$T_{dec}\simeq\frac{M}{\ln(m_{pl}/M)}.$$
The uniton density at the present universe is
$$Y_0=\frac{3.79 x_{dec}}{\sqrt{g_*} m_{pl} M <\sigma|v|>},$$
where we set $g_* =g_{*s}$, since the decoupling temperature is high.
Since the relic energy density of a massive decoupled particle is
$\rho_0=M s_0 Y_0$,
we can estimate the ratio of energy density to the critical energy density
at the present universe to be
$$\Omega_0 h^2 \equiv \frac{\rho h^2}{\rho_c}
\simeq 10^9 \frac{\ln(m_{pl}/M) M^2}
{N\alpha_s^2\sqrt{g_*}m_{pl}}{\rm GeV}^{-1}.$$
The cosmological data indicates that $0.1 <\Omega h^2<1 $.
Using this condition we get an upper bound on the mass of the uniton
\begin{equation}
M <10^5 \alpha_s \left(N\sqrt{g_*}
\ln(m_{pl}/M)\right)^{1/2}\mbox{GeV}\approx10^5~{\rm GeV}.
\end{equation}
If we assume the presence of inflation the bound on the uniton mass
is modified. In the case of inflation and with $T_R<T_{dec}$, the uniton
is diluted and is regenerated after reheating by out--of--equilibrium
production.
Since the uniton is completely diluted after the inflation,
the relic density at the reheating temperature is 0.
We can approximate it as
$$\frac{dY}{dx} =  \lambda x^{-2}Y_{eq}^2 \label{boltzman2},$$
with $Y_{eq}= 0.145 g/g_* x^{3/2} e^{-x}$.
Integrating this relation from the reheating temperature to the present
temperature
we get
$$Y_0= \frac{\lambda g^2}{2}
\left(\frac{0.145}{g_*}\right)^2
\left(x_r+\frac{1}{2}\right) e^{-2x_r},$$
where $x_r\equiv M/T_R$. $T_R$ is the reheating temperature, and
$$\Omega_0h^2 \simeq 9\times 10^3 N\alpha_s^2 g^2\frac{m_{pl}}{M}
\left(\frac{200}{g_*}\right)^{1.5}
\left(x_r+\frac{1}{2}\right) e^{-2x_r}.$$
We can estimate the bound on the mass,
\begin{equation}
M>T_R\left[25+\ln(\sqrt{M/T_R})\right].
\end{equation}
In this case the bound on the uniton mass depends on $T_R$.
It is noted that there are three windows (in the parameter space
$M/\sigma_p$, with $\sigma_p$ denoting the scattering cross section
on protons)
for strongly interacting dark
matter (such as $U_0$) \cite{SGED90}
which possibly meet our requirements.
The first window is
in the relatively low mass range  (10 GeV $< M < 10^4$ GeV )
and in the range $10^{-24}<\sigma_p< 10^{-20}$cm$^2$.
In other two windows it is required to have $10^5$ GeV $< M <10^7$ GeV
and $M>10^{10}$ GeV, respectively,
assuming a cross section, in both cases, less than
$10^{-25} $cm$^2$.
These constraints include bounds from various experiments
(such as experiments performed using solid state cosmic-ray
detectors and plastic track cosmic-ray detectors)
and from cosmological consideration
(such as the galactic halo infall rate and the life-time
of neutron stars \cite{neutron}).

The uniton is only one example of the exotic
Wilsonian matter states in the string
derived models. There are several other examples
with different properties. For example the models
contain color triplets which are weak singlets with
weak hypercharge $\pm1/6$. These color triplets
form bound mesonic and hadronic
states with the regular up and down quarks
which have fractional charge in multiples of $\pm1/2$.
At the same time the same models also contain
weak doublets and singlets with fractional
electric charge $\pm1/2$. Thus, these fractionally
charged baryons and leptons will form neutral bound
states, and the binding energy depends on their masses.
The evolution of these states in the early universe
is similar to that of the uniton since the three
gauge couplings are of approximately the same order.
The existence, however, of both baryons and leptons
with fractional electric charge $\pm1/2$ offers new
scenarios that may weaken the existing limits on
fractional charged matter. This is indeed an exciting
possibility that merits further investigation.

Another example of a Wilsonian matter state is the
state which is a Standard Model singlet with
``fractional'' charge under the $U(1)_{Z^\prime}$.
This state is similar to the right handed neutrino
but has half the $U(1)_{Z^\prime}$ of the right handed
neutrino. This type of states appear only in the
superstring derived standard--like models. For example,
in the model of refs. \cite{gcu} such states
appear from the sectors $b_{1,2}+b_3+\beta\pm\gamma$.
In this model these states transform as $3$ and ${\bar 3}$ of
a hidden $SU(3)_H$ gauge group. Their
interactions with the Standard Model states vanish to all
orders of nonrenormalizable terms if the $SU(3)_H$ is left unbroken.
Thus, these states interacts with the Standard Model states
only by the exchange of the $U(1)_{Z^\prime}$ vector boson.
The $U(1)_{Z^\prime}$ gauge symmetry has to be broken
somewhere in between the weak scale and the Planck scale.
Thus, depending on the scale of $U(1)_{Z^\prime}$ symmetry
breaking there are several possible dark matter scenarios
for the Wilsonian singlet.

\bigskip
1. $M>>M_{Z^\prime}$ without inflation. In this case
the Wilsonian singlet can annihilate into the Standard Model
fermions and into their superpartners, and into the $Z^\prime$
vector boson and its superpartner. The calculation is similar to the
corresponding calculation for the uniton and we obtain a similar limit.
$$M\le 10^{5}~\hbox{GeV}$$

\bigskip
2. $M>>M_{Z^\prime}$ with inflation and $T_R>M_{Z^\prime}$.
This case again is similar to the case of the uniton with inflation
and again we obtain a similar limit
$$M> T_R\left[25+{{1}\over{2}}\ln
\left({{M}\over{T_R}}\right)\right].$$

\bigskip
3. $M<<M_{Z^\prime}$ without inflation
$\rightarrow$ relativistic at decoupling

In this case the Wilsonian singlet is a WIMP and it can only
annihilate into the Standard Model fermions and their superpartners
via the $U(1)_{Z^\prime}$ interactions which are suppressed by
$1/{M_Z^\prime}$. Decoupling occurs when the Wilsonian singlet
is still relativistic. The number density in the comoving
volume is then estimated to be
$$Y_0\equiv \frac{n_{_{EQ}}}{{\rm s}}=
0.278\frac{g_{\rm eff}}{g_{*s(T_{\rm dec})}}
\simeq 1.2\times10^{-3}.$$
where the particle content of the MSSM is assumed and $T_{dec}>1~{\rm TeV}$.
We then obtain,
$$M<3~\hbox{keV}$$

\bigskip
4. $M<<M_{Z^\prime}$ with inflation.

In this case the Wilsonian singlet can be heavy. Inflation will
dilute the Wilsonian singlet and they will be regenerated
after reheating. There are two regions to consider, $T_R<M$ and $T_R> M$.
In the first case the $W_s$ decouples when it is non--relativistic.
In this limit we obtain a bound on the mass of the Wilsonian singlet
which is similar to the previous bounds in an inflationary scenario,
$$M> T_R\left[25+\frac{1}{2}\ln\left(\frac{M^5}
{M_{Z^\prime}^4T_R}\right)\right], ~~~~~~~~~~~~~~~~T_R<M.$$
In the relativistic limit ($T_R> M$) the cross section
is temperature dependent. In this case we approximate the
thermal average of the energy square by
$$\langle{s}\rangle=4\langle E^2 \rangle
\simeq\left[\frac{5}{4}\right]40 T^2.$$
and the cross section is given by
is given by
\begin{eqnarray}
\sigma|v| &\simeq& \frac{8}{3} N_{Z^\prime}
\pi \frac{{s}}{M^4_{Z^\prime}},~~~~~~~~~~~ {\rm if} ~T_R>M.
\label{go2}
\end{eqnarray}
Again we integrate the Boltzmann equation for the number density in
comoving volume from the reheating temperature to the present universe.
In this case we obtain a relation between three unknown parameters,
($M,\, M_{Z'},\, T_R$),
$$M<~{{M^4_{Z^\prime}\over{T_R^3}}6.9\times 10^{-25}
\left({{g_*}\over{200}}\right)^{1.5}
{{1}\over{N_{Z^\prime} g^2_{\rm eff}}}}
, ~~~~~~~~~~~~~~~~T_R>M.$$
\bigskip

Next, I turn to discuss the Wilsonian matter states in the
context of a superstring motivated dynamical SUSY breaking
scenario \cite{dsb}. In the dynamical SUSY breaking scenarios,
supersymmetry breaking is generated dynamically at a relatively
low scale and is transmitted to the observable sector by the
gauge interactions of the Standard Model \cite{dsbothers}.
Supersymmetry, in these scenarios,
is broken nonperturbatively in a hidden sector
and the breaking is mediated to the observable
sector by a messenger sector.
The universality of the Standard Model
gauge interactions results in generation blind mass parameters
for the supersymmetric scalar spectrum.
Consequently, in these scenarios supersymmetric
flavor changing neutral currents are naturally suppressed.
A crucial assumption in this regard is the absence of
interaction terms between the messenger sector
and the Standard Model states.

However, as seen above this is precisely what happens
in the case of the Wilsonian matter states. Namely,
the ``fractional'' charges of the ``Wilsonian matter states
result in unbroken local discrete symmetries which
forbid the coupling of the Wilsonian states to the
Standard Model states. Thus, the Wilsonian matter
states are natural candidates for the messenger
sector in the dynamical SUSY breaking scenarios.

A superstring motivated dynamical SUSY breaking scenario
was recently proposed \cite{dsb} in the model of ref. \cite{gcu}.
In this model the NS sector produces the generators of the
$SU(5)_H\times SU(3)_H\times U(1)^2$ hidden gauge group.
The hidden gauge group contains two non--Abelian factors, $SU(5)_H$
and $SU(3)_H$. In this model it is shown \cite{dsb} that the
requirement of a phenomenologically acceptable generation
mixing requires that $SU(5)_H$ is broken near the Planck scale
while $SU(3)_H$ is left unbroken. In this case the nonperturbative
interactions in the hidden $SU(3)$ gauge group may indeed be generated
at a relatively low scale, $\Lambda_3\approx100$ TeV, in accordance
with the low--energy gauge--mediated dynamical SUSY breaking scenarios.
As argued in ref.\ \cite{fh} a non--vanishing $F$--term may be generated
in the direction of one of the gauge singlets, $\xi_i$,
due to the hidden matter condensates. The analysis there
was done for the model of ref. \cite{eu}. However, because of
the similarities between the models I assume that
a similar $F$--term can be generated in this model as well.
The model of ref. \cite{gcu} contains two pairs of color triplets
of the uniton type, $\{D_1,~{\bar D}_1,~D_2,~{\bar D}_2\}$.
In the superpotential we find the couplings
$\xi D_1{\bar D}_1+\xi D_2{\bar D}_2$.
As shown above these uniton states do not have superpotential terms
with the Standard Model states. Therefore, this uniton states
are natural candidates for the messenger sector in the dynamical SUSY
breaking scenarios.

The problem of superstring gauge coupling unification motivates a
predictive hypothesis with regard to the messenger sector.
Dynamical SUSY breaking scenarios in the context of the
MSSM require the existence of both color triplets and electroweak
doublets in order not to spoil the unification of the gauge couplings.
It requires that the messenger sector states fall into complete
representations of $SU(5)$ and that the color triplets and
electroweak doublets are almost degenerate in mass.
These constraints makes the dynamical SUSY breaking scenarios in
the context of the MSSM somewhat ad hoc and unattractive.
However, in the context of the string derived models,
extra color triplets and doublets are in fact required
to obtain unification of the gauge couplings at the
string unification scale rather than at the MSSM unification scale.
Thus, in the context of the string models the existence of the
messenger sector is well motivated. Furthermore, in general,
the mass scale of the color triplets has to be much lighter than the mass
scale of the electroweak doublets. This motivates the hypothesis that
the messenger sector consists solely of color triplets.
Moreover, as shown previously the uniton dark matter scenario
requires that the uniton mass is of the order $M\approx100$ TeV,
which is precisely the mass scale which is required for the
uniton to play the role of the messenger sector in dynamical
SUSY breaking scenarios. The string of lucky strikes does not
end here. For if we assume the existence of color triplets as well
as electroweak doublets, then the charged lepton and the color triplets
are almost degenerate in mass. In this case both the color triplets
and the charged lepton are stable. While the color triplets, as was argued
above, can confine to form neutral bound states, the charged leptons
cannot. Stable charged leptons may be in contradiction with the
existence of neutron stars \cite{neutron}. This string of lucky
coincidences may be more than an accident, and motivates the hypothesis
that the messenger sector consists solely of color triplets.

The hypothesis that the messenger sector consists solely of
color triplets results in very specific predictions for
the supersymmetric spectrum. In the dynamical SUSY breaking
scenarios the gaugino masses are obtained by one--loop
exchange of the messenger sector states and are given by,
\begin{equation}
M_i(\Lambda)=c_i{{\alpha_i(\Lambda)}\over {4\pi}}\Lambda
\label{gauginomasses}
\end{equation}
where $\Lambda$ is the SUSY breaking scale, $c_i$ are coefficients which
depend on the messenger sector, and $\alpha_i(\Lambda)$ are
the Standard Model coupling constants at the scale $\Lambda$.
The scalar masses arise from two--loop diagrams and are given by
\begin{equation}
m^2(\Lambda)=
2\Lambda^2\left\{C_3\left[{{\alpha_3(\Lambda)}\over{4\pi}}\right]^2
                +C_2\left[{{\alpha_2(\Lambda)}\over{4\pi}}\right]^2
                +{3\over5}\left({Y\over2}\right)^2
                             \left[{{\alpha_1(\Lambda)}\over{4\pi}}\right]^2
\right\}
\label{scalarmasses}
\end{equation}
where the weak hypercharge has the standard $SO(10)$ normalization
$U(1)_Y=3/5 U(1)_1$ and $C_3=4/3$ for color triplet scalars and zero
for sleptons and $C_2=3/4$ for electroweak doublets and zero for singlets.

With the hypothesis that the messenger sector consists only of
color triplets $$M_2\equiv0.$$ The chargino mass matrix
is given by
\begin{equation}
M_{\tilde C}=\left(\matrix{{\tilde
M}_2&M_W\sqrt{2}\sin\beta\cr
                             M_W\sqrt{2}\cos\beta&\mu\cr}\right)~,
\label{chargino}
\end{equation}
With this hypothesis ${\tilde M}_2$ in eq. (\ref{chargino}) is
equal to zero and $\beta$ and $\mu$ are taken as free parameters.
Thus, in this scenario one of the eigenvalues of the chargino mass
matrix is predicted to be below the $W$--boson mass. Imposing the
current experimental limits on the supersymmetric spectrum
\cite{lepexp}, the lightest
chargino mass is predicted to be in the range
$$M_{\chi^\pm}\approx56-65~{\rm GeV}.$$
Similarly, from eq. (\ref{scalarmasses}) it is seen that in this
case the sneutrino is the lightest superparticle.

To conclude the Wilsonian matter states have important cosmological
and phenomenological implication. They give rise to natural dark matter
candidates whose stability is protected by a local discrete symmetry.
This is an important advantage of the Wilsonian dark over other
dark matter candidates whose stability relies on the
existence of global symmetries.
The Wilsonian matter states in the string derived models
offer exciting possibilities for confronting string inspired scenarios
with experimental data. For example, the hypothesis of dynamical
SUSY breaking with color triplets solely predicts a light chargino
and will be tested at LEP2. The existence of stable Wilsonian
matter states may be tested in dark matter searches and
in searches for rare elements.

\bigskip
{\bf acknowledgments:} I would like to thank Sanghyeon Chang and Claudio
Coriano for very fruitful and enjoyable collaboration on ref. \cite{ccf}
and I. Antoniadis, K. Babu, S. Dimopoulos, S. Thomas, C. Wagner, and
J. Wells for very helpful discussions on dynamical SUSY breaking.
I would like also to thank the Aspen Center For Theoretical Physics
for hospitality while part of this paper was written.
This work is supported in part by DOE Grant No.\ DE-FG-0586ER40272.

%=========================================================================
%======================== REFERENCES =====================================
%=========================================================================

\vfill\eject

\bigskip
\medskip

\bibliographystyle{unsrt}

\vfill\eject
\end{document}